\documentstyle[aaspp4]{article}
\def \sun {$_{\scriptscriptstyle \odot}$}
\lefthead{B\"ottcher \& Fryer}
\righthead{GRB Iron Lines}
\begin{document}
\begin{center} Submitted to {\em The Astrophysical Journal} on May 24, 2000
\par (revised: Sept. 14, 2000)
\end{center}
\vspace{1.cm}
\title{X-ray spectral features from GRBs:\\
Predictions of progenitor models}
\author{Markus B\"ottcher$^*$\footnote{Chandra Fellow} and 
Chris L. Fryer$^{\dagger}$\footnote{Feynman Fellow}}
\affil{$^*$Space Physics and Astronomy Department; Rice University, MS 108 \\
6100 S. Main Street; Houston, TX 77005 - 1892; USA \\
$^{\dagger}$Theoretical Astrophysics, Los Alamos National Laboratory;
Los Alamos, NM 87545; USA}
\authoremail{mboett@spacsun.rice.edu, clf@t6-serv.lanl.gov}

\begin{abstract}
We investigate the potentially observable prompt or delayed 
X-ray spectral features from the currently popular gamma-ray 
burst (GRB) models.  During the evolution of many GRB progenitors, 
a disk around the central GRB source is produced.  Shock heating 
as the GRB ejecta collide with the disk may produce observable 
X-ray features.  We first summarize predictions deduced from 
previous calculations which invoke photoionization and relativistic 
blast waves.  We then calculate the quasi-thermal X-ray line features 
produced assuming the ejecta are nonrelativistic (which is 
more likely for the disk interactions of many GRB models).
In the framework of the Hypernova/Collapsar model, delayed
(a few days -- several months after the GRB) bursts of 
line-dominated, thermal X-ray emission may be expected.
The He-merger scenario predicts similar X-ray emission 
line bursts $\lesssim$~a few days after the GRB. These X-ray 
signatures should be observable with {\it Chandra} and 
{\it XMM-Newton} out to at least $z \sim 1$. Weak emission 
line features $\lesssim$~a few days after the GRB may also 
result from the supranova GRB scenario. In all three cases,
significant X-ray absorption features, in particular during
the prompt GRB phase, are expected. No significant X-ray
spectral features should result from compact-object binary 
mergers.
\end{abstract}


\section{Introduction}

With the advent of the new generation of X-ray telescopes,
such as {\it Chandra} and {\it XMM-Newton}, the detection 
of X-ray spectral signatures from the environments of 
cosmological gamma-ray bursts (GRBs) has become a realistic 
prospect. The marginal detection of a redshifted Fe~K$\alpha$ 
emission line in the afterglow of GRB~970508 (\cite{piro99})
with the {\it BeppoSAX} NFI has stimulated a vital discussion 
about the possible origin of this line feature (\cite{ghi99};
\cite{lcg99}; \cite{boe99a}; B\"ottcher, Dermer, \& Liang 
\markcite{boe99b}1999b; \cite{vietri99}; \cite{weth00}; \cite{paerels00};
\cite{boe00}). So far, this discussion has concentrated on 
determining the required / inferred physical conditions in 
the vicinity of GRB~970508 needed to explain the iron-line 
feature in the afterglow (assuming this feature is real).  

An essential assumption in these papers was that the material
responsible for potential emission line features is illuminated
(and partially or completely ionized) by burst emission which 
is qualitatively similar to the observed GRB and afterglow radiation.  
Also, these papers assume that the blast wave which interacts 
with the line-emitting material is relativistic, and qualitatively 
similar to the one which is associated with the GRB and its afterglow.  
The conditions inferred under these assumptions are rather extreme, 
requiring large amounts ($\sim 10^{-4} \, M_{\odot}$) of iron to be 
concentrated anisotropically in a small ($R \lesssim 10^{-3}$~pc) region 
around the central engine of the GRB (e.g., \cite{lcg99}, \cite{weth00}, 
\cite{boe00}). Paerels et al. (\markcite{paerels00}2000) have
recently suggested that high-resolution X-ray spectroscopy may
help to distinguish between different line production mechanisms
as a way to unveal the underlying physical scenario. They
argue that, in particular, photoionization can be distinguished 
from collisional excitation of the line by virtue of the lower
plasma temperature resulting in photoionization scenarios. In their 
detailed analysis of the BeppoSAX MECS spectrum of GRB~970508
during the time segment in which the $\sim 3.6$~keV emission
line was marginally detected, they found that the line width
and apparent strength of the radiative recombination continuum
blueward of the emission line, in combination with the measured
redshift of the GRB at $z = 0.835$, are inconsistent with a
photoionization scenario in which the source of the line emission
is in photoionization equilibrium with the afterglow radiation.

In addition to this evidence, most of the currently popular 
gamma-ray burst models may be hard-pressed to produce the 
conditions inferred from photoionization-based scenarios to explain 
the observed iron-line feature in GRB~970508. These models are 
not ruled out, both because the iron line feature in GRB~970508 
is marginal, and even if it is real, it may appear only in one 
small subclass of GRBs. However, Vietri et al. (\markcite{vietri99}1999) 
have used this line to argue  for their alternative supranova model 
(\cite{sv98}) which invokes the collapse of a neutron star as it spins 
down. Unfortunately for this model, realistic calculations of collapsing 
neutron stars (Ruffert, Janka, \& Sch\"afer \markcite{ruf96}1996; Fryer 
\& Woosley \markcite{fry98}1998) find that these collapses eject too 
much baryonic material and have too little energy to produce GRBs.  
Because these models do not include magnetic fields, this does 
not rule out the supranova model.  Even so, the supranova model
would most likely produce a short burst, and hence 
can not explain the line feature seen in GRB~970508, since that 
burst was a long-duration burst with $t_{\gamma} \sim 25$~s. 

Instead of trying to explain GRB~970508, in this paper we look at 
the more viable ``black-hole accretion disk'' class of GRB engines 
(see Fryer, Woosley, \& Hartmann \markcite{fry99}1999 for a review) and 
estimate potential X-ray spectral features in GRB afterglows that 
these models produce.  We make the initial assumption that these 
features are produced by the collision of the GRB ejecta with the 
environment produced during the formation process of the GRB engine. 
In the Collapsar and he-merger engines, the GRB is surrounded by a 
disk of material which is produced during their progenitor evolution. 
MacFadyen \& Woosley (\markcite{mac99}1999) found that collapsar 
(and probably he-merger) GRBs are beamed, forming a jet along the 
angular momentum axis of the accreting disk.  They also found that 
the explosion along the equator is likely to be very baryon loaded 
($>$1\,M\sun) moving at velocities much less than the speed of light 
(less than $\sim 10^9$\,cm\,s$^{-1}$).  In \S 2, we discuss the 
formation of these nonrelativistic disks around black-hole accretion 
disk GRBs, estimating the structure of each disk.  We then 
use these structures in \S\S 3 and 4 to predict the X-ray afterglow 
signature of these GRBs. We conclude with a discussion of 
the implications these results have on current and future 
X-ray missions in \S 5.

We note that the predictions concerning potentially time-variable 
absorption features as derived by Ghisellini et al. (\markcite{ghi99}1999)
and B\"ottcher et al. (\markcite{boe99a}1999a) are valid, independent
of the detailed GRB mechanism. Obviously, material responsible for 
such X-ray absorption features is located along the line of sight to 
the burst, so that the observed GRB and afterglow radiation and the
assumption of a relativistic blast wave may be used.

\section{Disk formation}

Nearly all of the formation scenarios of black hole accretion 
disk GRBs invoke a ``common-envelope'' phase.  Binaries are 
said to be in a common-envelope phase when the hydrogen envelope 
of one star engulfs its companion (this usually occurs 
when the star expands into a giant or supergiant).  Friction and/or 
tidal forces cause the companion to spiral in towards the 
giant's helium core, ejecting the hydrogen envelope.  The 
evolution of two stars that enter a common envelope phase 
is one of the major uncertainties in binary population synthesis 
and, despite years of effort, still remains an open question 
(see Sandquist et al. \markcite{san98}1998 and references therein).  
This limits our ability to make any strong quantitative predictions 
of this phase, but we can use the latest simulations to guide our 
estimates.  Current simulations (e.g. Sandquist et al. \markcite{san98}1998) 
suggest that the companion inspiral occurs on a timescale of 1 -- 10 
orbital periods ($t_{\rm orbit}$).  The outcome of this inspiral is 
either:  a) a close binary system if the companion is able to eject 
the hydrogen envelope before it merges with the star's helium 
core or b) a merged object if the stars merge before the 
ejection of the hydrogen envelope.

The hydrogen envelope carries away much of the orbital angular 
momentum and is preferentially ejected in the orbital plane 
(\cite{san98}).  It is this material which forms the disks 
around GRB engines.  Double NS, BH-NS, and BH-WD mergers go 
through a common envelope phase long before they actually 
merge, and the progenitors of these GRBs travel far from their 
formation sites (and their common-envelope disks) before producing 
GRBs (e.g. \cite{fry99}, \cite{bb99} and references therein).  
Collapsar and he-merger GRBs, on the other hand, occur shortly 
after their common-envelope phase.  In this paper, we study the 
interaction of the ejecta from the GRB explosion with the 
common-envelope disks produced by collapsar and he-merger 
progenitors and determine the spectral features of this 
interaction.  But first we must estimate the structure and 
size of these disks.

\subsection{He-merger disks}

To understand the characteristics of the disks produced in he-mergers, 
we must first understand the formation process of he-mergers (for more
details, see \cite{fry99}). The progenitor of a he-merger GRB is a 
binary system with two massive stars (both stars have masses in excess 
of $\sim 8$~M\sun).  The more-massive star (primary) evolves through 
its life, collapsing to form a compact remnant (either a neutron star 
or black hole).  During the initial expansion, the primary may transfer 
mass to its companion (secondary star), and this mass transfer (or even 
a common-envelope phase) may tighten the orbital separation of these 
binaries.  For some binaries, asymmetries in the primary's 
supernova explosion also lead to a tighter binary system as the 
compact remnant is ``kicked'' into a closer orbit with its companion.  
In addition, the ejecta from the supernova explosion may enrich the 
envelope of the secondary (see \cite{isr99}) with r-process elements.

When the secondary evolves off the main sequence, it envelops the 
primary's compact remnant, and the binary goes into a common-envelope 
phase.  The compact remnant ejects the hydrogen envelope in a 
disk-like structure, but not before merging with the secondary's 
helium core.  A he-merger occurs after the compact remnant 
has settled into the helium core.  The inspiral process spins 
up the helium core, producing a disk around the compact remnant.  
In addition, the compact remnant accretes $\sim 1$ -- 3~M\sun during 
the inspiral and disk formation process (\cite{zha00}), causing it 
to collapse to a black hole if it is not one already.  This 
black-hole accretion disk system is surrounded by a disk 
formed from the enriched hydrogen envelope of the secondary.

There are no simulations of the common-envelope evolution of a 
compact remnant into a massive star which reliably predict the 
ejecta from the inspiral process.  However, we may extrapolate 
from simulations such as Sandquist et al. (\markcite{san98}1998) 
to obtain a rough estimate of the characteristics of the hydrogen 
disk. The ejection velocity ($v_{\rm ejection}$) is roughly the 
escape velocity at any given radius in the star:
\begin{equation}
\label{eq:vej}
v_{\rm ejection}=\sqrt{\frac{2 G M(r)_{\rm secondary}}{r}}.
\end{equation}
The time between GRB outburst and the ejection of matter is equal 
to the inspiral time (since the outburst occurs as soon as the 
neutron star spirals into the center of the secondary):
\begin{equation}
\label{eq:tgrb}
t_{\rm GRB} \approx 10 t_{\rm orbit} = 
\frac{20 \pi r^{1.5}}{G^{0.5} M(r)_{\rm secondary}^{0.5}}.
\end{equation}
At the time of the GRB outburst, the location of any layer of star with 
radius $r$ is simply:
\begin{equation}
\label{eq:Dej}
D_{\rm ejecta} \approx v_{\rm ejection} \times t_{\rm GRB} = 
20\sqrt{2} \pi r.
\end{equation}

\begin{figure}
\plotfiddle{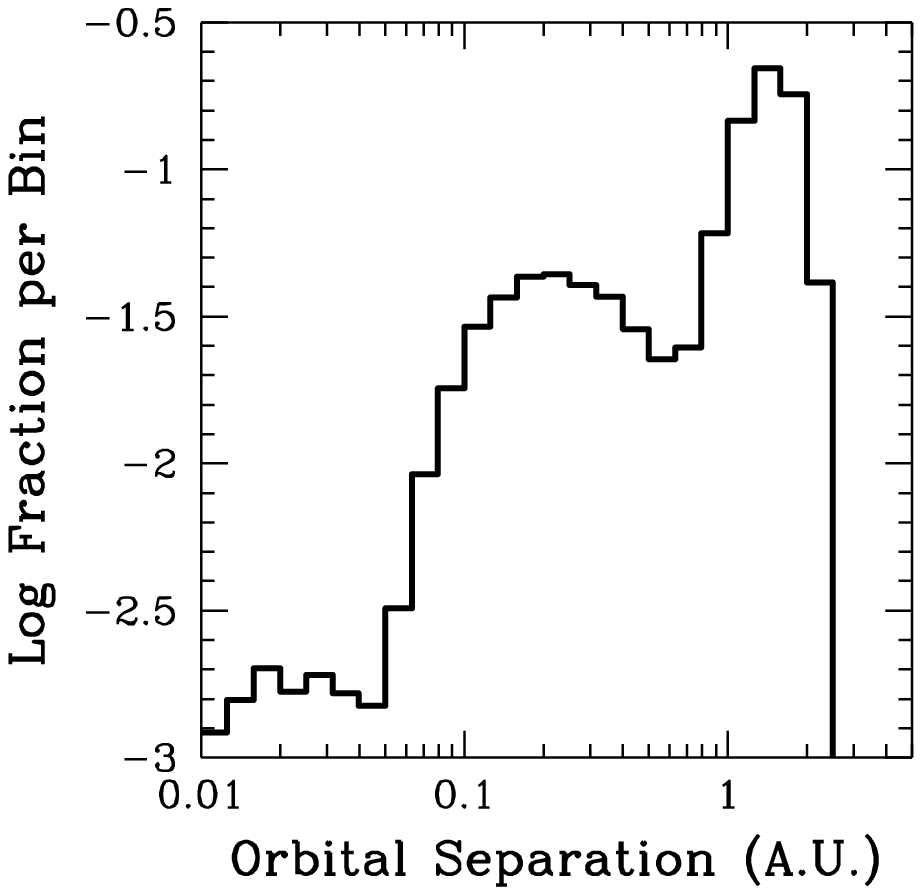}{7in}{0}{100}{100}{-170}{-50}
\caption{Distribution of orbital separations for he-mergers 
just prior to the common envelope phase, plotted as the 
fraction per bin (orbital separation is logarithmically spaced).
There are two peaks, one at 0.2 astronomical units, and the 
other at roughly 1.5 astronomical units.}
\label{fig:radii}
\end{figure}

\begin{figure}
\plotfiddle{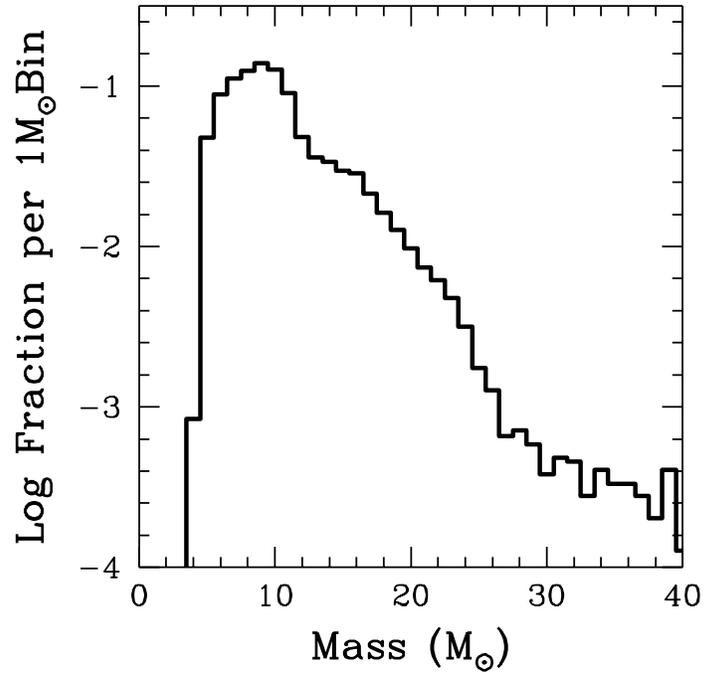}{7in}{0}{100}{100}{-170}{-50}
\caption{Distribution of companion masses for helium 
mergers.  Although the peak is at 9\,M\sun, it is likely 
that the more massive companions will produce more luminous 
bursts (their larger helium cores provide more fuel for 
the GRB).}
\label{fig:mass}
\end{figure}

\begin{figure}
\plotfiddle{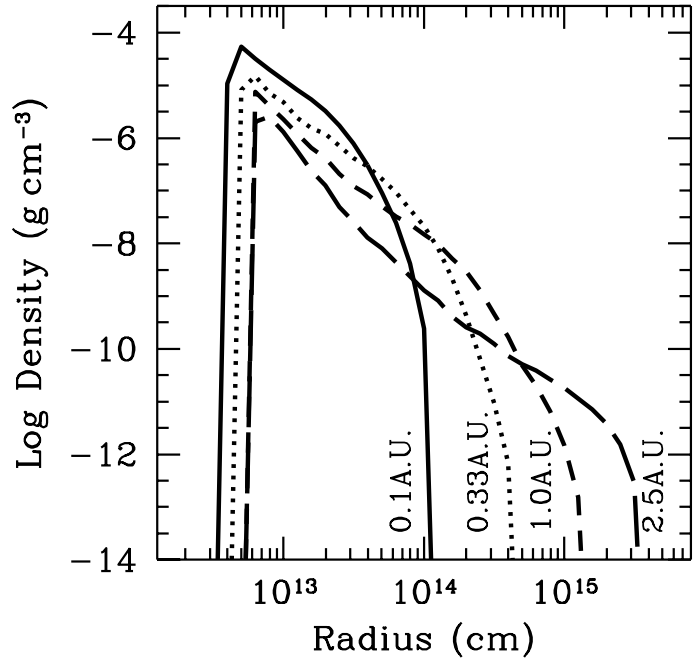}{7in}{0}{100}{100}{-170}{0}
\caption{Density vs. radius of the hydrogen disk formed during 
inspiral of a neutron star into a 15\,M\sun star (Heger 1999) as 
it expands off the main sequence for a range of binary separations:  
0.1, 0.33, 1, 2.5\,A.U.  We assume the star engulfs its compact 
companion when its radius exceeds the orbital separation.}
\label{fig:hedisks}
\end{figure}

\begin{figure}
\plotfiddle{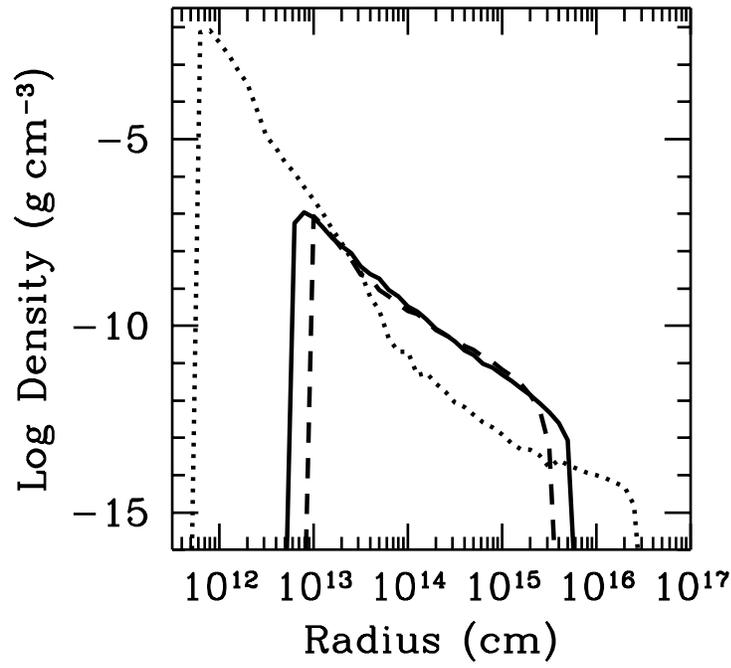}{7in}{0}{100}{100}{-170}{0}
\caption{Density vs. radius of the hydrogen disk formed during 
inspiral of a neutron star into 15, 25, and 40\,M\sun pre-collapse
stars (Woosley \& Weaver 1995).  The density profiles of these 
pre-collapse models give a good estimate of the structure of 
these stars after helium ignition (Case C mass transfer).}
\label{fig:he_masses}
\end{figure}

Clearly, the density structure of any he-merger disk depends upon 
the density structure of the secondary at the start of the 
common envelope phase.  This structure is a function of both 
the size of the star (and hence, the orbital separation of the 
binary) at the beginning of the common-envelope phase and the 
mass distribution of the companion star.  Using the binary population 
synthesis code developed in Fryer et al. (\markcite{fry99}1999), 
we can calculate the distribution of orbital separations and 
companion masses for he-merger progenitors (Figs.~\ref{fig:radii} 
and \ref{fig:mass}). Fig.~\ref{fig:hedisks} shows the range of 
hydrogen disk structures formed by he-mergers with a 15~M\sun companion 
star for a series of orbital separations. Note that the outer 
disk radius is most sensitive to the orbital separation prior to 
the common envelope evolution.  Fig.~\ref{fig:he_masses} shows the 
range of disk structures formed by 15, 25, and 40~M\sun stars 
assuming the orbital separation is set to the maximum radius of the 
companion.  Although our rough estimates of the disk formation and  
the uncertainties in binary population synthesis and stellar evolution 
(see \cite{fry99} for a discussion) make it difficult to predict the 
density profile of these disks accurately, these figures give a 
flavor of the range of possible disk structures.

\subsection{Collapsar disks}

The ring around Supernova 1987A proves that at least some massive 
stars have disks.  The progenitor of Supernova 1987A was probably 
a binary system in which the more massive primary engulfed its 
companion, causing the companion to spiral into the primary  
during a common-envelope phase (\cite{pod92}). The companion was 
unable to eject the entire hydrogen envelope and it merged with the 
primary's helium core.  This process could lead to the formation 
of an outward moving disk which, in the case of supernova 1987A, 
was lit up 10,000 -- 100,000 years later by the supernova to reveal 
a ``ring'' (see for example, \cite{col99}).
 
The majority of Collapsar progenitors also follow an evolutionary 
path where a binary system goes through a common envelope phase 
although the companion often does not merge with the primary
(\cite{fry99}).  The ``classical'' Collapsar model requires a 
massive helium star (without a hydrogen envelope) to avoid baryon 
contamination (\cite{mac99}), and a common-envelope phase is required 
to eject most of the hydrogen envelope (\cite{fry99}). Unlike 
the he-merger disks, the common envelope phase can occur 
more than 100,000 years before the eventual GRB outburst.  Although 
the Collapsar progenitor is still likely to be surrounded by this 
ejecta, disks formed in Collapsar progenitors will be much 
further away at outburst than those disks formed in he-mergers.
Assuming an ejection velocity equal to the escape velocity 
(eq. \ref{eq:vej}) and setting the time from disk ejection to GRB 
outburst ($t_{\rm GRB}$) to 100,000~y, we find that the inner edge 
of most Collapsar disks exceeds $10^{17}$~cm.  

However, if the binary does not go into a common-envelope until just 
before the collapse of the primary, a much more compact disk may 
be formed.  Recall that a common-envelope phase occurs when the 
radius of a star expands enough to engulf its companion.  Most 
stellar models reveal that a star actually contracts during the 
last 10,000 -- 100,000~y of its life (\cite{sch93}, \cite{ww95}).  
If the binary does not enter a common-envelope phase before 
this contraction, it is unlikely that it ever will.  So with 
the current stellar models, $t_{\rm GRB}$ must be greater than 
10,000 -- 100,000~y.  However, stellar models do not accurately 
predict the radii of massive stars, and there is a growing concern 
that these radii may be drastically incorrect (\cite{fry99}, 
\cite{wel99}, \cite{fry00}).  It is possible that massive stars 
reach their maximum size just before collapse.  If the common-envelope 
phase occurs 10 -- 100 years before collapse, the inner edge of the 
disk could be less than $10^{15}$~cm.  Until accurate stellar models 
are produced, we can not refine our estimates further.  However, even 
with these rough estimates of the disk structure, we can now estimate 
the expected X-ray afterglow spectral features from both of these 
objects.

X-rays may also be produced when the GRB ejecta strikes the 
Collapsar's companion star.  Recall that the companion generally 
does not merge with the collapsar during the common envelope 
phase.  The ejecta will hit this companion star, causing it 
to heat and expand, producing an X-ray emitting nebula 
(P. Pinto - private communication).  The magnitude and spectra 
of X-rays under this mechanism is difficult to predict 
quantitatively, and we will delay further discussion of this 
emission for a later paper.

\section{Predicted X-ray spectral features}

\subsection{Analytic estimate of the maximum iron line luminosity}

In this section, we present a simple analytic estimate for an upper 
limit to the total, isotropic luminosity in the two major constituents 
of the iron K$\alpha$ line blend from a hot, highly ionized plasma, 
namely the Fe~XXV He$\alpha$ (2p1s $\to$ 1s$^2$ resonance) and the
Fe~XXVI H$\alpha$ (2p $\to$ 1s) transitions. Apart from energy 
conservation constraints, the luminosity in the resonant emission 
lines considered here is restricted by several line destruction
mechanisms, such as Compton scattering, photoelectric absorption
by lighter elements, and collisional de-excitation. Considering
Compton scattering and photoelectric absorption --- both of which
processes would remove a line photon from the line --- one can 
define an effectively emitting volume given by the fraction 
of the total disk volume through which the optical depth
$\tau_{\rm L} \equiv \tau_{\rm T} + \tau_{\rm pa} \equiv 
\tau_{\rm T} \, (1 + f_{\rm pa})$ due to electron scattering 
and photoelectric absorption is $\sim 1$. In a neutral plasma of
cosmic element abundance, we have $f_{\rm pa} \sim 4$ at the energies
of the $n = 2 \to 1$ transitions of Fe~XXV and Fe XXVI. In an ionized
plasma, this number obviously becomes smaller, and for our simple
estimate we assume $f_{\rm pa} \sim 2$. The effect of collisional
de-excitation becomes relevant if the density of the line-emitting
material becomes comparable to or greater than the critical density
$n_{\rm crit}$ at which the collisional de-excitation rate equals
the spontaneous decay rate. Using van Regemorter's (\cite{vanrege62})
$\overline g$ approximation, the critical density can be roughly
approximated as 

\begin{equation}
n_{\rm crit}^{\rm c.d.} \approx 2.4 \times 10^{11} {\sqrt{T_e / {\rm K}} 
\over \lambda_{\rm nm}^{3}} \; {\rm cm}^{-3} \approx 3.8 \times 10^{17}
T_8^{1/2} \; {\rm cm}^{-3},
\label{ncrit}
\end{equation}
where $T_8 = T / (10^8 \, {\rm K})$.

For the ease of computation, we assume that the emitting region 
can be geometrically represented by a torus, located at a distance 
$r = 10^{13} \, r_{13}$~cm from the center of the burst source, 
with a cross-sectional radius of $a = 10^{12} \, a_{12}$~cm, 
containing a total mass of $M_{\rm T} = m_{\rm T} \, M_{\odot}$, 
the volume of the torus will be given by $V_{\rm T} = 2 \times 10^{38} 
\, a_{12}^2 \, r_{13}$~cm$^3$. The average density of the torus 
material is then $n_{\rm T} = 6 \times 10^{18} \, [ m_{\rm T} / 
(a_{12}^2 \, r_{13})] \; {\rm cm}^{-3}$. If the observer is looking 
along a line of sight close to the symmetry axis of the torus, then 
the Thomson depth through the torus is reasonably well approximated 
by

\begin{equation}
\tau_{\rm T} = a \, n_{\rm T} \, \sigma_{\rm T} = 4 \times 10^6 
\left( {m_{\rm T} \over a_{12} \, r_{13}} \right).
\label{tau_T}
\end{equation}

We define a critical density for Thomson scattering, 
$n_{\rm crit}^{\rm T}$ as the density at which the Thomson 
depth equals 1, so that for densities $n > n_{\rm crit}^{\rm T}$ 
line photons are likely to be scattered out of the line before 
leaving the emitting volume. This critical density is given 
by $n_{\rm crit}^{\rm T} \approx 1.5 \cdot 10^{12} \, 
a_{12}^{-1}$~cm$^{-3}$. In the situations of interest here,
we always find $n_{\rm crit}^{\rm T} \ll n_{\rm crit}^{\rm c.d.}$,
indicating that line destruction by Compton scattering and
photoelectric absorption is dominant over collisional de-excitation.

The luminosity in the emission lines may thus be estimated as 
$L_{\rm line} = j_{\rm L} \, V_{\rm T} / \max\left\lbrace 
\tau_{\rm L}, \, 1 \right\rbrace$, where $j_L$ is the emissivity 
in the line. In order to parametrize the line emissivities by an 
emissivity parameter $x$, we use the notation of Raymond \& Smith 
(\markcite{rs77}1977), where $j_{\rm L} = n_{\rm H} \, n_e \cdot 
10^{- 23 - x}$~erg~cm$^{-3}$~s$^{-1}$. The emissivity parameters 
are inferred from runs of the XSTAR code (\cite{kmc82}) with a 
negligibly small ionization parameter of 
$\xi = 10^{-8}$~ergs~cm~s$^{-1}$ and constant, pre-specified 
plasma temperature. This yields

\begin{equation}
L_{\rm line} \approx 6 \cdot 10^{45 - x} \, {m_{\rm T} \over a_{12}} 
\; {\rm ergs \; s}^{-1}.
\label{est_lum}
\end{equation}

\begin{figure}
\plotfiddle{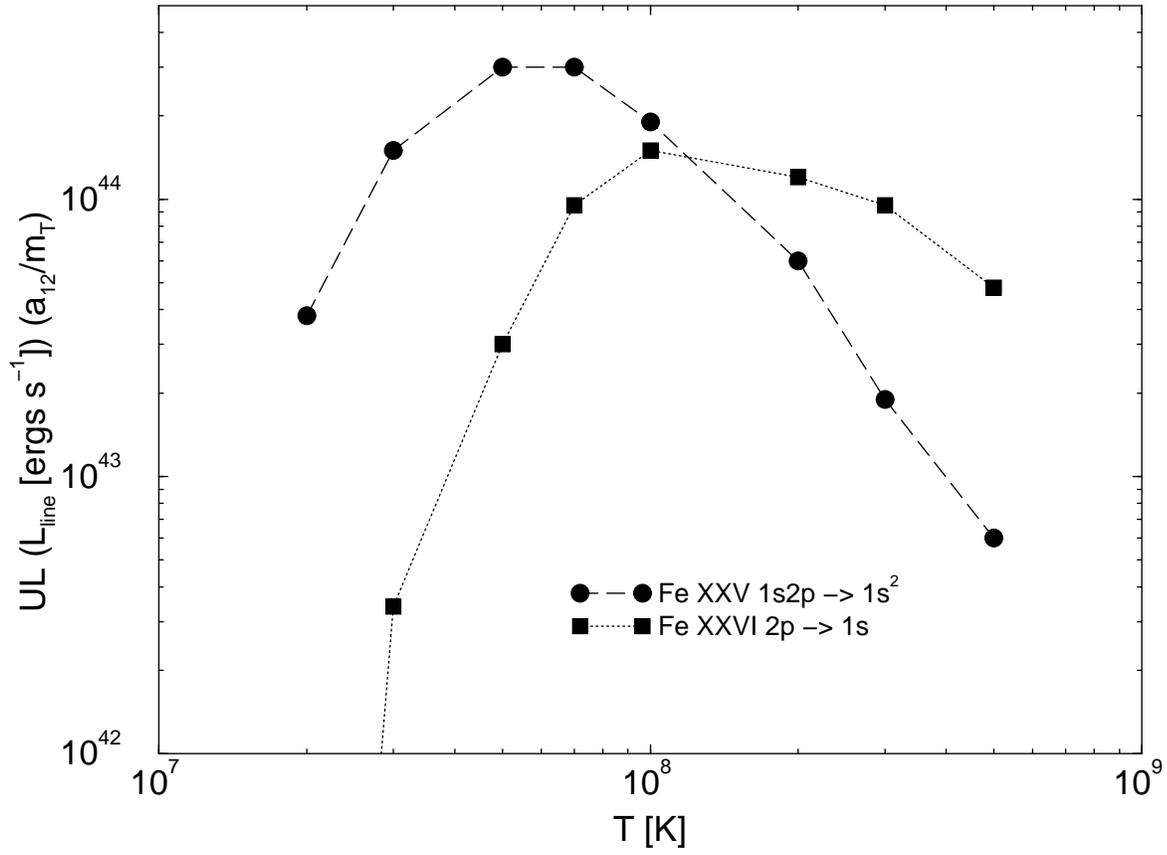}{7in}{-90}{70}{70}{-280}{500}
\caption{Estimated upper limits on the luminosity in the Fe~XXV~He$\alpha$
($1s2s \, \to \, 1s^2$) and Fe~XXVI~H$\alpha$ ($2p \, \to \, 1s$) lines at
$\sim 6.7$~keV from a hot, thermal torus at $r = 10^{13} \, r_{13}$~cm, as 
a function of temperature of the torus material.}
\label{ul_graph}
\end{figure}

Some representative values of the emissivity parameters $x$ and the 
upper limits on the line luminosities are listed in Table 
\ref{upper_limits}, and the estimated luminosity upper limits 
are plotted as a function of plasma temperature in 
Fig.~\ref{ul_graph}. Although these are very crude estimates, 
they indicate that Fe~K$\alpha$ luminosities in excess of $\sim 
10^{44}$~ergs~s$^{-1}$ are very well possible in a shock-heated-disk 
scenario.

\subsection{Simulations of the shockwave / disk interaction}

In order to get a realistic prediction of the expected X-ray line
and continuum emission from the shock-heated disk, we simulate the
shock-wave evolution and thermal history of the shocked material
as the supernova ejecta, associated with the GRB explosion, 
interact with the pre-ejected disk. We assume that in the course 
of the supernova/GRB a total mass of $M_0 \sim 1 \, M_{\odot}$
is ejected quasi-isotropically at a speed of $v_s = 10^9 \, 
v_9$~cm~s$^{-1}$, initiating a shock-wave when interacting with 
the disk of pre-ejected material. 

For the case of an intermediate-mass (secondary) progenitor 
(15, 25~$M_{\odot}$), we parametrize the density 
profile of the disk by $\rho_{\rm d} (r) = \rho_{\rm 0} \, (r 
/ r_{\rm in})^{-2.5}$ for $r_{\rm in} \le r \le r_{\rm out}$ 
and its geometry by a constant $h/r \sim 0.1$. For more massive 
stars ($40 \, M_{\odot}$) the density profile may be approximated 
by a broken power-law with $\rho_{\rm d} (r) = \rho_{\rm 0} \, 
(r / r_{\rm in})^{-4}$ for $r_{\rm in} \le r \le r_{\rm br}$, and
$\rho_{\rm d} (r) = \rho_{\rm 0} \, (r_{\rm br} / r_{\rm in})^{-4}
\, (r / r_{\rm br})^{-1}$ for $r_{\rm br} \le r \le r_{\rm out}$
(see Fig. \ref{fig:he_masses}). Writing $r_{\rm in} = 10^x \, 
r_{i, x}$~cm and $n_0 = \rho_0 / (\overline A \, m_p) = 10^{17} 
\, n_{17}$~cm$^{-3}$, where $\overline A$ is the average atomic 
weight of the disk material, the total mass in the pre-ejected 
disk is

\begin{equation}
M_{\rm d} = 0.1 \, \left( {h/r \over 0.1} \right) \, n_{17} \, 
r_{i, 13}^3 \, \left(\sqrt{r_{\rm out} \over r_{\rm in}} - 1 
\right) \; M_{\odot}
\label{Md25}
\end{equation}
for intermediate-mass progenitors, and
\begin{equation}
M_{\rm d} = 0.5 \, \left( {h/r \over 0.1} \right) \, n_{21} \, 
r_{i, 12}^3 \left( 1 + {r_{\rm in} \, r_{\rm out}^2 \over
2 \, r_{\rm br}^3} \right) \; M_{\odot}
\label{Md40}
\end{equation}
for high-mass progenitors.

The deceleration of the non-relativistic shock wave and the
heating of the ejecta and and the swept-up material, are calculated
by numerically solving simultaneously for the energy and momentum
equations,

\begin{equation}
{d \over dt} \left( M \, v_{\rm s} \right) = 4 \pi \, R_{\rm s}^2 \, 
\overline P_{\rm s},
\label{momentum}
\end{equation}
\begin{equation}
{dM \over dt} = 4 \pi R_{\rm s}^2 \, \rho_{\rm d} (R_s) \, v_{\rm s},
\label{mass}
\end{equation}
\begin{equation}
{dE \over dt} = - \overline P_{\rm s} \, {dV \over dt} + \dot E_{\rm rad},
\label{dEdt}
\end{equation}
where
\begin{equation}
E = {1 \over 2} \overline \rho_{\rm s} \, v_{\rm s}^2 \, V +
{\overline P_{\rm s} \, V \over \gamma - 1},
\label{total_energy}
\end{equation}
$R_{\rm S}$ is the shock radius, $\overline P_{\rm s}$ 
and $\overline\rho_{\rm s}$ are the volume-averaged pressure 
and density of the shocked material, $V$ is the volume occupied 
by the shocked material, $\gamma$ is its adiabatic index, and 
$\dot E_{\rm rad}$ is the radiative cooling term. Throughout 
this paper, we assume $\gamma = 5/3$. In Eq. \ref{momentum}, 
we have neglected the pressure of the disk material.

Due to the high densities involved and to the fact that the shock wave 
is non-relativistic, we may assume that at any point in time the shocked 
material is in approximate thermal and collisional ionization equilibrium. 
Photoionization precursors are not expected to play an important role since 
the Compton scattering depth $\lambda_C = (n \, \sigma_{\rm T})^{-1}
\approx 1.5 \times 10^9 \, n_{15}^{-1}$~cm is much smaller than any 
characteristic size scale of the system, so that most of the fluorescence 
photons produced in the photoionization precursor will be absorbed within 
the disk and thus be unobservable. The emission from the shocked region
can therefore be represented by pure thermal plasma emission from an 
optically thick plasma at the temperature of the shocked material. At
any given time, we calculate the emission from the shocked material
using XSTAR (Kallman \& McCray \markcite{kmc82}1982) in a 
constant-temperature, purely thermal ionization mode (i. e. 
with very small photoionization parameter). 

\section{Results}

\begin{figure}
\plotfiddle{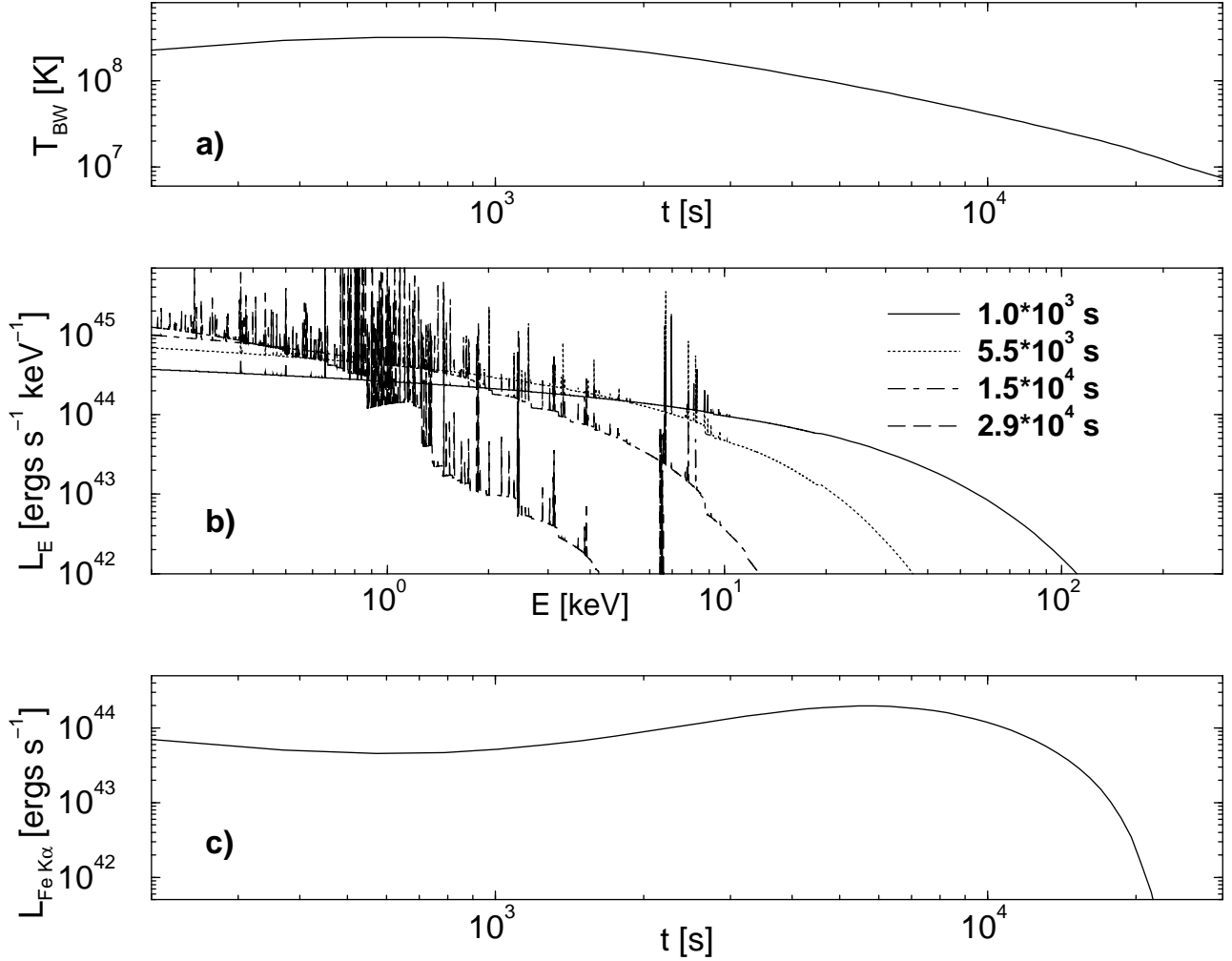}{7in}{-90}{70}{70}{-280}{500}
\caption{Temporal evolution of the temperature (top panel),
X-ray spectra (middle panel), and light curve of the emission
in the Fe~K$\alpha$ line (bottom panel) for $1 \, 
M_{\odot}$ ejected at $10^9$~cm~s$^{-1}$ and interacting
with the disk of pre-ejected material from a $25 \, M_{\odot}$
progenitor with disk inner radius $r_{\rm in} = 10^{13}$~cm. 
$t = 0$ corresponds to the time of the onset of the 
blast-wave/disk interaction, i. e. $(1 + z) \, r_{\rm in} / 
v_{\rm s} \sim (1 + z) \times 10^4$~s after the GRB/supernova 
explosion. All times and photon energies quoted in the figure 
refer to the cosmological rest frame of the GRB source.}
\label{r13_graph}
\end{figure}

We have done a series of simulations for a variety of parameters
representative of both he-merger and collapsar/hypernova disks.
In Fig. \ref{r13_graph} we show the temporal evolution of the
temperature of material behind the shock, resulting sample
X-ray spectra at different times after the onset of the shock
wave / disk interaction, and the light curve of the Fe~K$\alpha$
line luminosity, for a disk with inner radius at $r_{\rm in} =
10^{13}$~cm, representative for the he-merger case. In those 
simulations, we have assumed a disk density profile appropriate
for a $25\, M_{\odot}$ progenitor, and a mass of $1 \, M_{\odot}$
for the ejected material. The ejecta are assumed to have an initial 
velocity of $10^9$~cm~s$^{-1}$, corresponding to a kinetic energy of 
$10^{51}$~ergs in the ejecta. The figure shows that, especially in 
the later, decaying phase, $\gtrsim 2 \, (1 + z) \times 10^4$~s 
after the onset of the shock wave / disk interaction, the thermal 
X-ray spectrum from the shocked disk material is strongly line 
dominated and might yield excellent prospects of detection by 
X-ray telescopes sensitive at $\lesssim 1$~keV.

Tab. \ref{table_r13} illustrates how the maximum Fe~K$\alpha$ 
line luminosity and the decay time constant of the line emission 
depend on the mass and velocity of the ejecta shock-heating the 
pre-ejected disk from the $25 \, M_{\odot}$ progenitor. The time 
constant $t_{\rm d}$ is determined by fitting an 
$\exp[-(t/t_{\rm d})^2]$ law to the decaying portion of the 
iron line light curves. We find that the shock wave / disk
interaction expected for the he-merger scenario can very 
plausibly produce an Fe~K$\alpha$ line of apparent quasi-isotropic 
luminosity $L_{{\rm Fe \, K}\alpha} \sim 10^{44}$~ergs~s$^{-1}$ 
maintained over $t_{\rm d} \lesssim (1 + z) \times 10^4$~s after 
the ejecta begin to interact with the disk. 

\begin{figure}
\plotfiddle{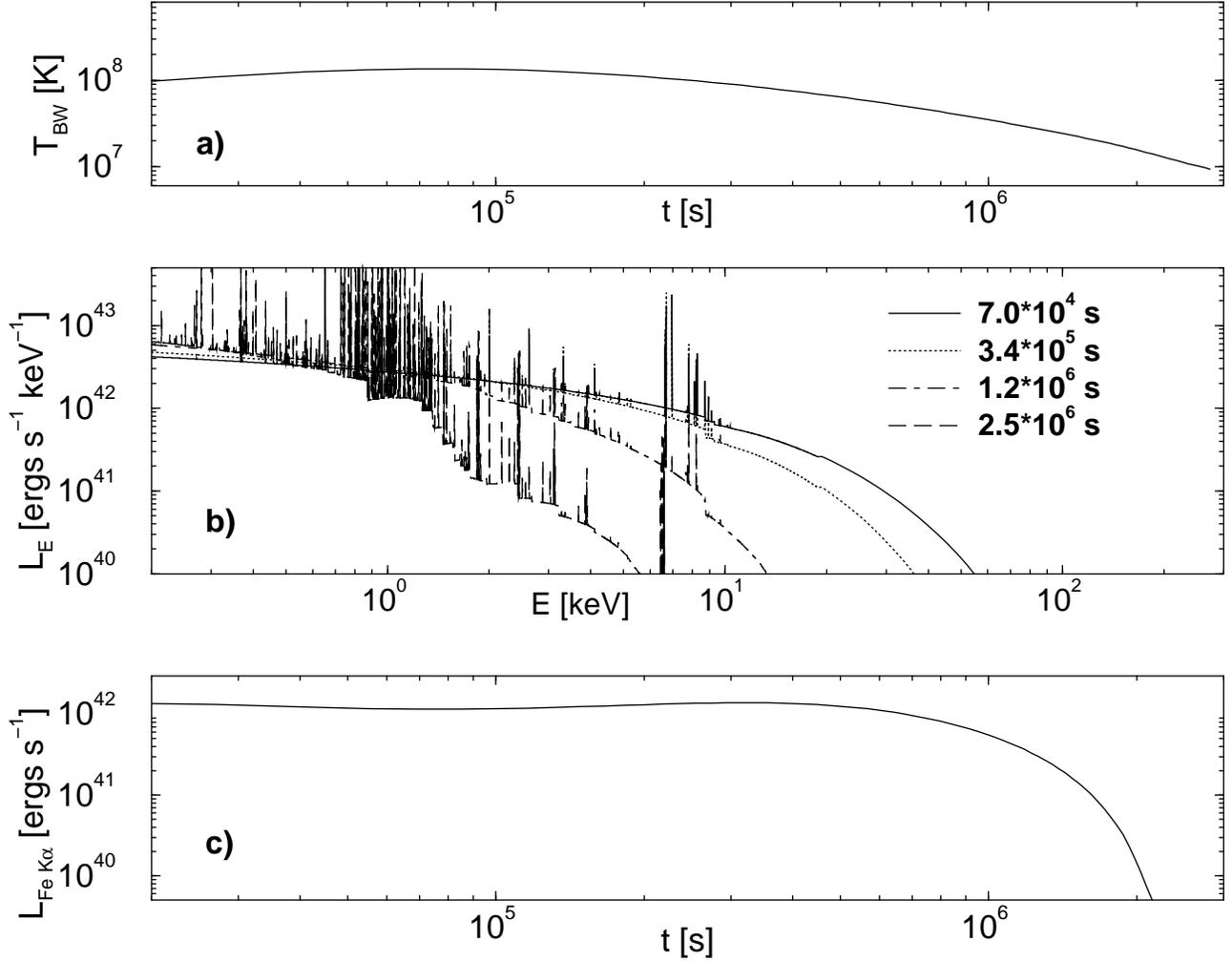}{7in}{-90}{70}{70}{-280}{500}
\caption{Same as Fig. \ref{r13_graph}, except for disk
inner radius, $r_{\rm in} = 10^{15}$~cm. Thus, here 
$t = 0$ corresponds to $\sim (1 + z) \times 10^6$~s after 
the GRB/supernova explosion.}
\label{r15_graph}
\end{figure}

Fig. \ref{r15_graph} shows the temperature evolution, X-ray spectra,
and Fe~K$\alpha$ light curve for a case representative of a collapsar
/ hypernova disk, if the system entered the common-envelope phase 
$\sim 100$~y before the GRB. The progenitor is assumed to be a 
$25 M_{\odot}$ star. In this case, the disk inner edge is expected
to be located at $r_{\rm in} \sim 10^{15}$~cm. Results of simulations
with different ejecta mass and velocity are summarized in Tab.
\ref{table_r15}. In this case, maximum iron line luminosities
in excess of $\sim 10^{42}$~ergs~s$^{-1}$ are still possible,
while the typical time delay between the GRB and the onset of
the GRB is now $\sim (1 + z) \times 10^6$~s.

Comparing simulations with identical disk mass and ejecta 
mass and velocity, but different inner disk radii, we find 
an approximate scaling law for the maximum Fe~K$\alpha$ line 
luminosity, $L_{{\rm Fe \, K}\alpha} \propto r_{\rm in}^{-1}$. 

\section{Observational prospects}

In the previous section, we found that in both the
helium-merger and in the collapsar/hypernova scenarios,
a quasi-thermal X-ray flash from the shock-heated disk may
result. A critical and yet very uncertain parameter (especially
for the collapsar/hypernova scenario) in the model simulations 
is the inner disk radius, which is primarily determined by the 
duration of the common-envelope phase prior to the GRB event. 
In the he-merger scenario, we expect typically $r_{\rm in} \sim 
10^{13}$~cm, while in the collapsar/hypernova scenario, this
parameter could have values $10^{14}$~cm~$\lesssim r_{\rm in} 
\lesssim 10^{17}$~cm. The onset and decay time scale of the
resulting secondary X-ray flash scale as $\Delta t_X \propto
r_{\rm in}$, while the peak X-ray luminosity is approximately
$L_{{\rm Fe \, K}\alpha} \propto r_{\rm in}^{-1}$. Since the
continuum X-ray afterglows from the (probably beamed) relativistic 
ejecta typically decay with temporal indices $\chi \gtrsim 1.2$, 
(if $F_X (t) \propto t^{-\chi}$), detection of the secondary, 
thermal X-ray outbursts predicted by the he-merger and 
collapsar/hypernova scenarios might be favored by larger 
disk radii, as long as the resulting X-ray flux remains 
above the detection threshold of currently operating X-ray 
telescopes. 

\begin{figure}
\plotfiddle{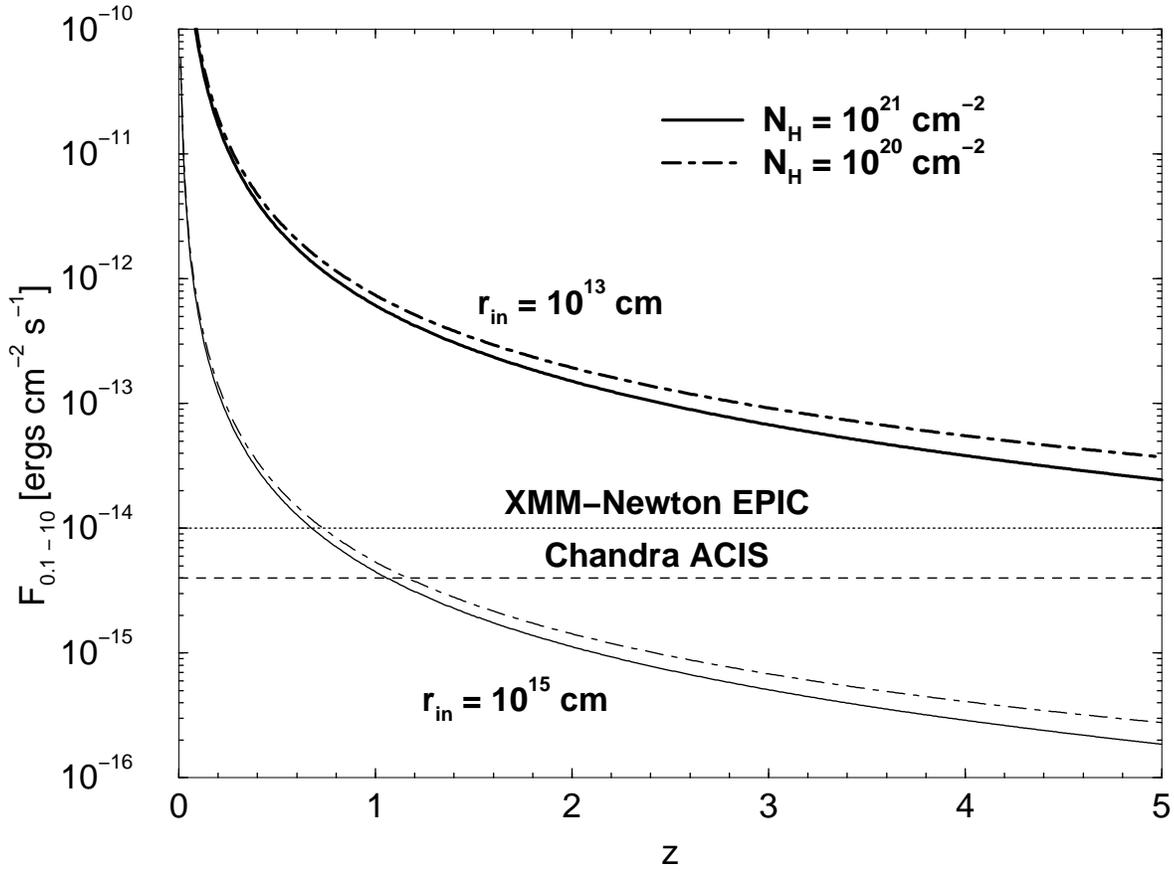}{7in}{-90}{70}{70}{-280}{500}
\caption{Peak 0.1 -- 10~keV X-ray fluxes, accounting for
Galactic absorption, as a function of redshift, for the
blast wave simulations illustrated in Fig. \ref{r13_graph}
(thick lines) and Fig. \ref{r15_graph} (thin lines).
Horizontal lines show the nominal, estimated sensitivity
limits of the {\it Chandra} ACIS and {\it XMM-Newton} EPIC
detectors, respectively, for a 10~ksec exposure. }
\label{F_z}
\end{figure}

Fig. \ref{F_z} shows the absorbed 0.1 -- 10~keV peak fluxes 
resulting from two of our simulations, as a function of redshift
of the GRB source. Shown are representative cases for $r_{\rm in}
= 10^{13}$~cm and $r_{\rm in} = 10^{15}$~cm (see also Figs.
\ref{r13_graph} and \ref{r15_graph}), corresponding to onset 
delays of $\sim (1 + z) \times 10^4$~s and $\sim (1 + z) \times 10^6$~s,
respectively. Two plausible values of the Galactic neutral 
hydrogen column density, $N_H$, are used. Given that the nominal
point source sensitivity for a 10~ksec exposure on {\it Chandra}'s
ACIS detectors is $\sim 4 \times 10^{-15}$~ergs~cm$^{-2}$~s$^{-1}$,
while for the EPIC detectors on board {\it XMM-Newton} this limit
is $\sim 10^{-14}$~ergs~cm$^{-2}$~s$^{-1}$, the predicted X-ray 
flashes from the shock wave / disk interaction may be detectable 
out to redshifts of at least $z \sim 1$ for an inner disk radius 
of $r_{\rm in} \sim 10^{15}$~cm. For a source at $z = 1$ with
$r_{\rm in} = 10^{15}$~cm, we would expect the onset of the 
secondary X-ray burst $\sim 3$~weeks after the GRB. 

\section{Summary}

During the progenitor evolution of collapsar/hypernova and he-merger 
GRBs, a hydrogen disk is formed around the central engine. These GRB 
engines produce a jet along the disk axis, and the relativistic outburst 
which produces the gamma-rays does not interact with this disk.  
However, both of these engines are likely to expel $\gtrsim 1$~M\sun 
along the equator at lower velocities ($\sim 10^9$~cm~s$^{-1}$). 
The interaction of the explosion ejecta with the expelled disk 
may produce X-ray luminosities in excess of $\sim 10^{44} \, 
(r_{\rm in} / 10^{13} \, {\rm cm})^{-1}$~ergs~s$^{-1}$ with a 
delay of $\sim (r_{\rm in} / 10^{13} \, {\rm cm}) \, (1 + z) 
\times 10^4$~s after the GRB.  For gamma-ray bursts with a redshift 
$z \lesssim 1$, this emission is well within the capabilities of the 
latest X-ray satellites (e.g. {\it Chandra} and {\it XMM-Newton}).
Thus, long-term monitoring of X-ray afterglows over several weeks
after the GRB (most notably, even after the direct afterglow
radiation has faded to undetectable levels) may lead to the 
detection of these secondary X-ray flashes which would yield
valuable information about the nature and pre-burst evolution 
of the GRB progenitor.

\acknowledgements{The work of M.B. is supported by NASA through
Chandra Postdoctoral Fellowship Award Number PF~9-10007, issued 
by the Chandra X-ray Center, which is operated by the Smithsonian
Astrophysical Observatory for and on behalf of NASA under contract 
NAS~8-39073.  C.L.F was supported by a Feynman Fellowship at LANL, 
NSF (AST-97-31569), and the US DOE ASCI Program (W-7405-ENG-48).  
It is a pleasure to thank P. Pinto for helpful advice and 
discussions.}

\clearpage

\clearpage

\clearpage

\clearpage

\begin{deluxetable}{ccccccc}
\tablewidth{14cm}
\tablecaption{Emissivity parameters $x$, and upper limits on the
resulting luminosity in the $\sim 6.7$~keV iron lines of Fe~XXV 
and Fe~XXVI from a hot thermal torus.}
\tablehead{
\colhead{T [K]} & \colhead{$x_{{\rm He} \, \alpha}$} & 
\colhead{${{\rm UL} (L_{{\rm He} \, \alpha}) \, a_{12} \over m_{\rm T}}$ [ergs~s$^{-1}$]} &
\colhead{$x_{{\rm H} \, \alpha}$} & 
\colhead{${{\rm UL} (L_{{\rm H} \, \alpha}) \, a_{12} \over m_{\rm T}}$ [ergs~s$^{-1}$]} 
}
\startdata
$2 \times 10^7$ & 2.2 & $3.8 \times 10^{43}$ &
                  7.0 & $6.0 \times 10^{38}$ \nl
$3 \times 10^7$ & 1.6 & $1.5 \times 10^{44}$ &
                  3.4 & $2.4 \times 10^{42}$ \nl
$5 \times 10^7$ & 1.3 & $3.0 \times 10^{44}$ &
                  2.3 & $3.0 \times 10^{43}$ \nl
$7 \times 10^7$ & 1.3 & $3.0 \times 10^{44}$ &
                  1.8 & $9.5 \times 10^{43}$ \nl
$1 \times 10^8$ & 1.5 & $1.9 \times 10^{44}$ &
                  1.6 & $1.5 \times 10^{44}$ \nl
$2 \times 10^8$ & 2.0 & $6.0 \times 10^{43}$ &
                  1.7 & $1.2 \times 10^{44}$ \nl
$3 \times 10^8$ & 2.5 & $1.9 \times 10^{43}$ &
                  1.8 & $9.5 \times 10^{43}$ \nl
$5 \times 10^8$ & 3.0 & $6.0 \times 10^{42}$ &
                  2.1 & $4.8 \times 10^{43}$ \nl
\enddata
\label{upper_limits}
\end{deluxetable}

\begin{deluxetable}{ccccc}
\tablewidth{10cm}
\tablecaption{Simulated maximum quasi-isotropic Fe~K$\alpha$ line 
luminosities and decay time scales $t_{\rm d}$ for ejecta of different 
mass and velocities shock-heating the pre-ejected disk from a $25 \, 
M_{\odot}$ progenitor, $r_{\rm in} = 10^{13}$~cm.}
\tablehead{
\colhead{$v_{\rm ej}$} &
\colhead{$M_{\rm ej}$} & 
\colhead{$E_{\rm ej}$} &
\colhead{$L_{{\rm K}\alpha} (t_{\rm max})$} & 
\colhead{$t_{\rm d}$} \nl
\colhead{[$10^9$~cm~s$^{-1}$]} & 
\colhead{[$M_{\odot}$]} &
\colhead{[$10^{51}$~ergs]} &
\colhead{[$10^{43}$~ergs~s$^{-1}$]} & 
\colhead{[$10^3$~s]}
}
\startdata
1     & 0.5 & 0.5   & 14  & 11  \nl 
1     & 1   & 1     & 18  & 12  \nl 
1     & 2   & 2     & 29  & 13  \nl 
0.707 & 1   & 0.5   & 16  & 9.5 \nl 
1.414 & 1   & 2     & 20  & 16  \nl 
0.707 & 0.5 & 0.25  & 11  & 8.5 \nl 
0.5   & 0.5 & 0.125 & 9.1 & 7.0 \nl 
\enddata
\label{table_r13}
\end{deluxetable}

\begin{deluxetable}{ccccc}
\tablewidth{10cm}
\tablecaption{Simulated maximum quasi-isotropic Fe~K$\alpha$ line 
luminosities and decay time scales $t_{\rm d}$ for ejecta of different 
mass and velocities shock-heating the pre-ejected disk from a $25 \, 
M_{\odot}$ progenitor, $r_{\rm in} = 10^{15}$~cm.}
\tablehead{
\colhead{$v_{\rm ej}$} &
\colhead{$M_{\rm ej}$} & 
\colhead{$E_{\rm ej}$} &
\colhead{$L_{{\rm K}\alpha} (t_{\rm max})$} & 
\colhead{$t_{\rm d}$} \nl
\colhead{[$10^9$~cm~s$^{-1}$]} & 
\colhead{[$M_{\odot}$]} &
\colhead{[$10^{51}$~ergs]} &
\colhead{[$10^{41}$~ergs~s$^{-1}$]} & 
\colhead{[$10^5$~s]}
}
\startdata
1     & 0.5 & 0.5   & 8.7 & 8.0 \nl 
1     & 1   & 1     & 14  & 9.0 \nl 
1     & 2   & 2     & 27  & 9.0 \nl 
0.707 & 1   & 0.5   & 13  & 6.5 \nl 
1.414 & 1   & 2     & 13  & 12  \nl 
0.707 & 0.5 & 0.25  & 8.0 & 6.5 \nl 
\enddata
\label{table_r15}
\end{deluxetable}

\end{document}